\documentclass[conference]{llncs}
\usepackage{blindtext}
\usepackage{enumitem}
\usepackage{makecell}
\usepackage{lipsum}
\usepackage{url}
 \usepackage{capt-of}
\usepackage{graphicx}
\usepackage{tikz}
\usepackage{pgfplots}
\usetikzlibrary{arrows,positioning,automata}
\usepackage{balance}
\usepackage{amsmath}
\usepackage[T1]{fontenc}
\usepackage{algorithm}
\usepackage{algpseudocode}
\usepackage{algorithm}
\usepackage{subfigure}
\usepackage{graphicx}
\usepackage[utf8]{inputenc}
\usepackage{balance}
\usepackage{color}
\usepackage{comment}
\usepackage{hyperref}
\usepackage{array}
\usepackage{soul}
\usepackage{color}
\usepackage{lipsum}
\usepackage[justification=centering]{caption}
\usepackage{authblk}

\newcommand{\eat}[1]{}

\begin{document}
\title{Using Social Media for Word-of-Mouth Marketing}

\author{Nagendra Kumar \and Yash Chandarana\and K. Anand \and Manish Singh\vspace{-2ex}}
\institute{\textit{Indian Institute of Technology Hyderabad, India – 502285}\\
\textit{Email: \{cs14resch11005, cs14btech11040, cs14resch11004, msingh\}@iith.ac.in}}

\renewcommand\Authands{ and }

\maketitle

\begin{abstract}
Nowadays online social networks are used extensively for personal and commercial purposes. 
This widespread popularity makes them an ideal platform for advertisements. 
Social media can be used for both direct and word-of-mouth (WoM) marketing. 
Although WoM marketing is considered more effective and 
it requires less advertisement cost, it is currently being under-utilized. 
To do WoM marketing, we need to identify a set of people 
who can use their authoritative position in social network to promote a given product. 
In this paper, we show how to do WoM marketing in Facebook group, which is a
question answer type of social network. We also present concept of  
reinforced WoM marketing, where multiple authorities can together
promote a product to increase the effectiveness of marketing. 
We perform our experiments on Facebook group dataset 
consisting of 0.3 million messages and 10 million user reactions. 
\end{abstract}

\section{Introduction}
\label{sec:fipsn_int}
Marketing is a process by which products and services are introduced and promoted to potential customers. Marketing leads to increase in sales, build the reputation of company and maintain healthy competition. To do effective marketing, one has to identify the best customers, understand their needs and implement the most effective marketing method. There are many marketing methods such as direct marketing, field marketing, account-based marketing, B2P marketing, online marketing, word-of-mouth marketing, etc. With the emergence of Internet, online marketing has become one of the biggest sources of marketing. In online marketing, advertisements provide range from basic text descriptions with links to rich graphics with slideshows. However, the problem with most of these advertisement strategies is the lack of trust that users have on these information sources. People are being bombarded with so many online advertisements that they have grown immune to online advertisements. Word-of-Mouth~(WoM) marketing has the advantage that the advertisement is done by people who are trusted by the person whom we try to market. According to Whitler\footnote{http://www.forbes.com/sites/kimberlywhitler/2014/07/17/why-word-of-mouth-marketing-is-the-most-important-social-media/7762b8f07a77}, 64\% of marketing executives indicate that they believe WoM marketing is the most effective form of marketing. Incite~\cite{incite} also stated that 91\% of B2B (business-to-business) buyers are influenced by WoM marketers when making their buying decisions.
In WoM marketing the information is passed from person to person through the WoM communication. People believe on the words of people whom they know such as friends, family and closely known authorities.  If we do WoM marketing through only friends and family then the marketing will be quite restricted. Since people use multiple social media to access different types of information, we propose the use of social media to do widespread WoM marketing. 

There are different social network models, such as friend-to-friend, follower-following, question-answer, etc. In this paper, we use question-answer~(QA) type of network to do WoM marketing. QA network can lead to more widespread marketing compared to other types of networks because the influential users in such networks are known by more number of users compared to other types of networks. For example, in a friend-to-friend network a user may just have few hundred friends and the user may not even have an authoritative status amongst his peers. There are many QA networks, such as Quora, Stack Overflow, Facebook groups, etc. In this paper, we use online social groups (OSGs) such as Facebook groups. The members of a Facebook group have more focused interest compared to a generic friends or follower-following networks. For example, Java, Java for developers, C/C++ programming, etc., are some very popular and focused public groups having more than 20,000 group members. 

Since Facebook groups are focused on specific topic and have large number of members, one can use the prominent and reliable members of such groups to do marketing. Prominent members, whom we call influential members, post lots of important and relevant information. In Facebook groups, members can ask questions from other members of the group to get solution to their problems. Influential members help other members by posting useful information in the form of posts and comments, and in return they get publicity in the form of reactions, such as likes, comments, shares, from other members. Organizations can use these trusted influential users to market their products by giving them incentives.
Since the recommendations are made by one of their trusted peers, with influential position, members of group pay more attention to such recommendations. Let's consider an example task to better appreciate our problem.

\vspace{-0.1in}
\begin{example} 
A book publisher wants to advertise a book, say a DBMS book, in Facebook group. It has a limited advertisement budget. It would like to find few influential users who can promote the book to a large audience. The publisher can attract such influential users by giving some free sample copy or discount. Following are some questions that would be of interest to the publisher:
\vspace{-0.05in}
\begin{enumerate}  
\item For a given Facebook group, who are the top-$k$ influential users?
\item What fraction of the group would be influenced by a selected set of top influential users?
\item How to do reinforced marketing so that each topic is marketed jointly by at least $k$ influential users?
\item What is the best time to start promotion in the group? 
\end{enumerate}
\end{example} 
\vspace{-0.05in}
In this paper, we answer all the above questions. Our key contributions are as follows:
\begin{itemize}
\item We use Facebook groups for product marketing. 
\item We analyze the characteristics of Facebook groups. 
\item We propose different marketing strategies and give solution using social network analysis.
\item We present a topical relevance method to create social interaction graph from the users' activities in the group. 
\item We find important characteristics of influential users in Facebook groups and examine the dynamics of user influence across topics and time. 
\item We evaluate our algorithms on a large dataset containing 0.3 million posts and 10 million reactions. 
\end{itemize}

\section{Related Work}
\label{sec:fipsn_rw}
The subject of social media and network marketing has attracted significant research attention~\cite{domingos2001mining,trusov2010determining}. Trusov et al.~\cite{trusov2010determining} stated that social networking firms earn from either showing advertisements to site visitors or being paid for each click/action taken by site visitors in response to an advertisement. Domingos et al.~\cite{domingos2001mining} studied the mining of the network value of customers. They have shown that network marketing which exploits the network value of customers, can be extremely effective. According to Ogilvy Cannes\footnote{http://www.adweek.com/prnewser/ogilvy-cannes-study-behold-the-power-of-word-of-mouth/95190?red=pr} 74\% of consumers identify WoM marketer as a key influencer in their purchasing decisions.

According to MarketShare~\cite{MarketShare}, WoM has been shown to improve marketing effectiveness up to 54\%. It has been shown by Wu et al.~\cite{wu2011says} that less than 1\% of the social network users produce 50\% percent of its content, while the others have much less influence and completely different social behavior~\cite{weng2010twitterrank}. However, in case of Facebook groups we observe that 6.5\% users generate 85\% content of the group and less than 2\% of these users are able to influence 80\% population of the group. These statistics show the users' behavior of Facebook groups and it is quite different from other social networks such as friend network in Facebook, Twitter, etc.

To discover top authorities in social network, we need to understand topological structure of social network, flow and diffusion of information in social network~\cite{cheng2014can,leskovec2007patterns,guille2013information}. Many researchers in this domain have studied the structure of social network to study the similar problem of influence maximization ~\cite{chen2009efficient,kempe2003maximizing}. The goal of Influence maximization is to maximize product penetration, while minimizing the promotion cost by selecting the subset of users which are also called influential users. Vogiatzis et al.~\cite{vogiatzis2013influential} stated that influential users could spread the news of the product or service may reach up to maximum possible level. However, our approach is complementary to the existing approaches of finding influential users. These works did not focus on finding topical influential users based on interaction activities. Our findings show that influence of a user varies across the topics and influential users may not be interested in chosen advertising product. Moreover, users’ friend network is small and most of them do not mention their interests. 
We find out the users' interests from their activities in social network groups which are focused communities and choose top users who are authoritative users as well as interested in the advertising the product.

\section{Problem Definition}
\label{sec:fipsn_plm}
We define the problem of finding the influential users for WoM marketing in terms of the following sequence of sub-problems:

\vspace{0.05in}
\noindent {\bf Problem 1 (Create social interaction graph)}: {\it Given a topic $T$, a Facebook group $F$ and the activities $A$ in the group $F$, create a social interaction graph $G(V,E)$, where the vertices $V$ represent the members of the group $F$ and edges $E$ represent the interaction between group members.}

Problem 1 is to create a social interaction graph for a given Facebook group. 
The activities in a group include creation of posts and reactions to posts, such as likes, comments, likes on comment, and shares. 
The members of the group represent vertices and interaction among members (users) represent edges of the graph. 
It is a topic sensitive graph, where the edge weights are dynamically computed based on the given topic $T$. 
We assign a weight to the edge based on the given topic $T$, type of reaction that a user had done to the post or comment created by an other user.

\vspace{0.05in}
\noindent {\bf Problem 2 (Finding influential users)}: {\it Given a social interaction graph $G(V,E)$ and a topic $T$, find the top-K influential users $I$ from the graph $G(V,E)$, who can give maximum visibility to the topic $T$ in the corresponding Facebook group of the given interaction graph $G(V,E)$.}

People form Facebook groups to explore about certain topic. Naturally, in such groups some members with more knowledge become authorities, whose words have great influence on the other group members. In this problem our goal is to find the influential users for a given marketing topic.

\vspace{0.05in}
\noindent {\bf Problem 3 (Reinforced marketing)}: {\it Given a social interaction graph $G(V,E)$, a topic $T$ and reinforcement parameter $r$, find the set of influential users $I_R$ such that the marketing of each influential user from $I_R$ can be reinforced by at least $(r-1)$ other influential users.}

The social position of a user has important effect on the marketing. If someone who is not an authority markets a product, the marketing will hardly have any impact. If one authority markets the product, the marketing will be more effective. The marketing will be even more effective if multiple authorities can collectively market the product. When people hear the same message reinforced by multiple authorities that they trust, it is more likely they will consider buying the product. Thus, we need to find authorities in such a way that if one authority markets the product, there are at least $(r-1)$ other authorities in the set $I_R$, who can support the marketing. These $(r-1)$ other authorities should be closely related with other members whom the above mentioned one authority will market.

\section{Analysis of Online Social Groups}
\label{sec:fipsn_niosggs}
\vspace{-0.1in}
In this section, we present structure of 
OSGs to get insight into the users' activities. 
We use bow tie structure~\cite{broder2000graph} to 
analyze the general structure of OSGs. 
It has five distinct components namely \textit{core}, \textit{in}, \textit{out}, \textit{tendrils} and \textit{tube}. 
In bow tie structure, \textit{core} is a strongly connected component and 
contains users who often help each other. 
The \textit{in} component contains users who only react to the posts. 
The \textit{out} consists of users who only post the contents. 
\textit{Tendrils} and \textit{tubes} contain the users who connect to either \textit{in} or \textit{out} or both but not to the \textit{core}. 
\textit{Tendrils} users only react to the posts created by \textit{out} users or 
whose posts are only reacted by \textit{in} users. 
\textit{Tubes} users connect to both \textit{in} and \textit{out}.

\vspace{-0.2in}
\begin{figure}
\centering
\begin{minipage}{.5\columnwidth}
  \centering
   \includegraphics[width=6cm, height=4cm]{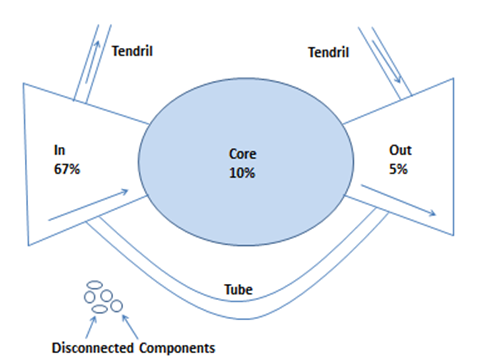}
  \captionof{figure}{Bow tie structure of the OSGs}
   \vspace{-0.01in}
  \label{fig:bowtie}
\end{minipage}%
\begin{minipage}{.5\columnwidth}
  \centering
  \includegraphics[width=6cm, height=4cm]{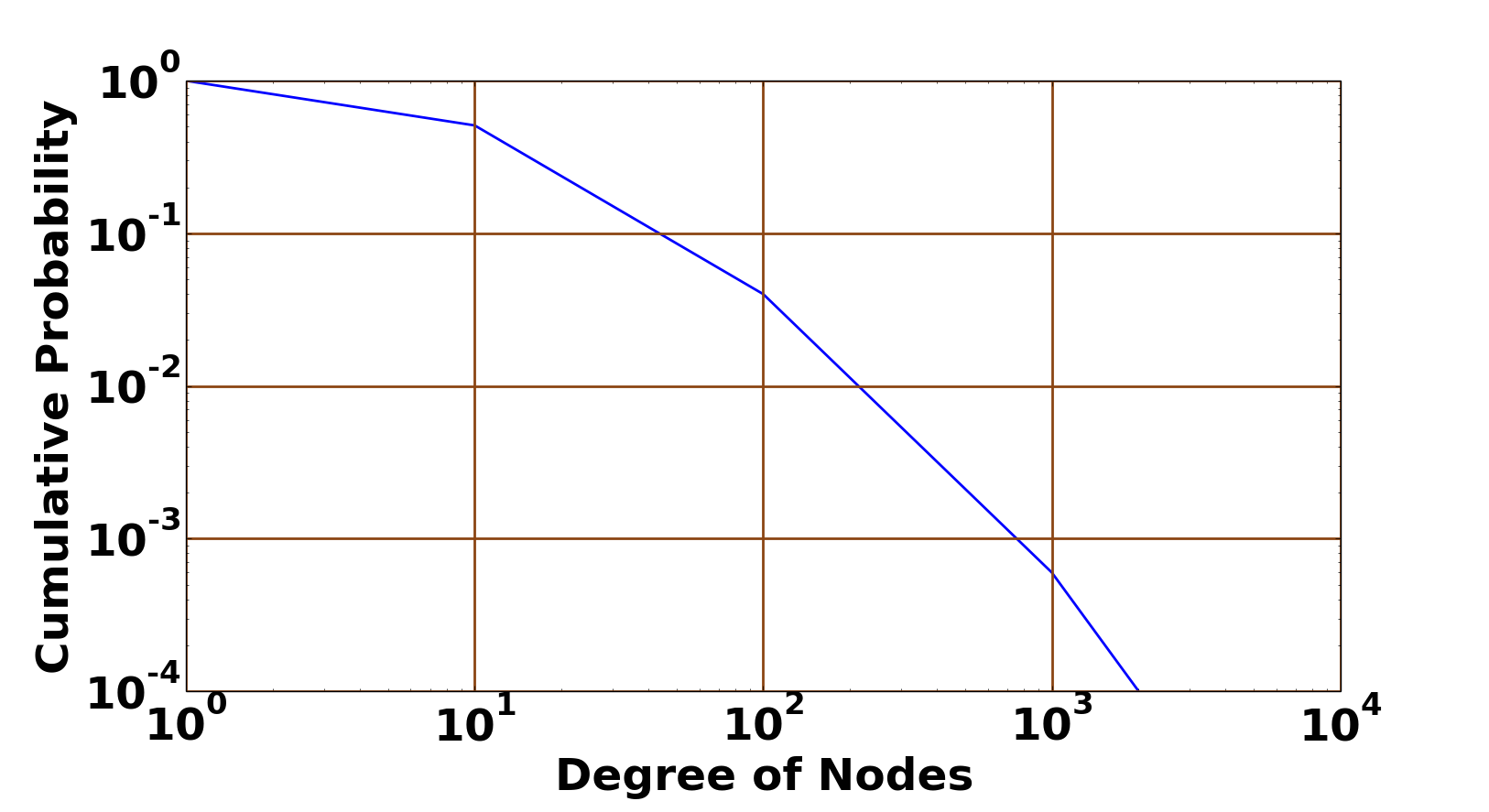}
  \captionof{figure}{Degree distribution in groups}
     \vspace{0.005in}
  \label{fig:degvsprb}
\end{minipage}
\end{figure}
 \vspace{-0.22in}

We present the bow tie structure of OSGs in Figure~\ref{fig:bowtie}. 
We observe that there are 10\% \textit{core} users, 67\% \textit{in} users, 5\% \textit{out} users, 15\% \textit{tendril} users, 0.2\% \textit{tube} users and 2.8\% \textit{disconnected} users. OSGs have much bigger \textit{in} component  compared to the \textit{out} and \textit{core}. We find that in OSGs about 10\% of the users do both, post contents and react to the posts of each other. Most of the users (67\%) only react to the posts, and 5\% of the users only post the contents. These results indicate that OSGs are the information seeking communities where most of the users consume the information and very few people generate the information. Most of the members join the groups to keep themselves updated by getting the information related to the topics of shared interest. 

Next, we perform degree analysis to get more insight into the users' connectivity in the groups. Degree is a general way to illustrate users’ relative connectedness in a large complex network. Degree distribution reports the number of users (cumulative probability of users) in the network with a given degree. 
As we can see in Figure~\ref{fig:degvsprb} that the degree distribution appears to follow a power law. 
The most of the users have very less degree which signifies that these users are connected to just a few other users; however, there are very few users who are connected to a large number of users. As these few users have large number of connections, they can easily spread  the information to a large audience.

\section{Social Interaction Graph}
\label{sec:fipsn_sig}
In this section, we give a solution to Problem 1. 
We generate the topic sensitive social interaction graph based on the group activities. 
We consider the topical relevance of group members to create the graph. 

\subsection{Measuring Topical Relevance}
We measure the topical relevance of users by analyzing the content of their posts. For a given topic $T$, we find the users who are interested in topic $T$. For example, to market a \textit{database} book, we need to find the users who have posted contents related to \textit{database}. One simple approach is to look into all the posts which contain the word \textit{database} in them. However, this approach fails to give good results as there might be posts which are actually relevant to the database but do not contain the word \textit{database}. In order to identify such posts, we need to generate a list of words which are semantically related to the given seed word. For example, for the word \textit{database}, some related words can be \textit{sql}, \textit{query}, \textit{schema}, etc. Clearly, the word \textit{sql} is more closely related to \textit{database} as compared to the word \textit{query}. For this task we need a system which, given a word, gives a list of relevant words along with it's relevance score. 

In this paper, we use Semantic Link\footnote{http://semantic-link.com/} system, which gives a list of words which are semantically related to the seed word. It uses the fact that some words occur frequently together. For example, the words \textit{database} and \textit{sql} often occur together. These are semantically related words, meaning that their co-occurrence is not due to chance but rather due to some non-trivial relationship. Semantic Link attempts to find such relationships between words and uses these relationships to find the related words. Semantic Link analyzes the text of the English Wikipedia and attempts to find all pairs of words which are semantically related. It uses a statistical measure called Mutual Information~(MI), which is a measure of the mutual dependence between two topics. Higher the MI score for a given pair of topics, higher the chance that they are related. MI score is defined as follows:
\begin{equation}
MI(x,y)= \sum_{y \in Y} \sum_{x \in X} p(x,y) \log{\dfrac{p(x,y)}{p(x)p(y)}}
\end{equation}
Where, $X$ and $Y$ are two random set of topics. $p(x,y)$ is the joint probability distribution function of $X$ and $Y$. p(x) and p(y) are the marginal probability density functions of $X$ and $Y$ respectively. 

After getting a list of related words, we find posts relevant to these words. One approach is to filter out the posts which do not contain any of the related words. However, this approach has a limitation that the users who don’t have relevant posts (related words in their posts) will have no in-links. Such users will get low ranks while applying authority measures on the graph and thus their out-links will not contribute much to the rank of the relevant users. In such a case, only the popularity of the relevant users will matter while determining the ranks. This approach ignores the relationship of relevant users with non-relevant users. A better approach is to give higher weight to the relevant posts and their interactions. The weight (boosted relevance) is calculated on the basis of the presence of relevant words in the posts. First, we calculate $relevance$ score for every post which is the sum of the MI score of all the related words that are present in the post. We then compute boosted relevance ($bRelevance$) based on the $relevance$ score. 
 \vspace{-0.1in}
  \begin{algorithm}
\caption{Algorithm for Computing Boost to Interactions}
\label{Algo:boost}
\begin{tabular}{ll}
    \textit{Input:} & $T$: set of topic words\\
    & $P$: Post\\
  \textit{Output:} & $bRelevance$: boost of post P\\
  \textit{Method:} & \\
    \end{tabular}
\begin{algorithmic}[1]
           \State $relevance \gets 0 $
           \State $postWords \gets P.getWords() $
	\ForAll{$tWord \in T$}
	\ForAll{$pWord \in postWords$}
	        \State $relevance+ = Similarity(tWord, pWord)$
        \EndFor
            \EndFor
        \State  $bRelevance = 1+\alpha*ln(1+relevance)$
         \State \textbf{return}  $bRelevance$
\end{algorithmic}
\end{algorithm}
\vspace{-0.1in}

Algorithm~\ref{Algo:boost} shows the method that we use to compute the $bRelevance$ of different posts. Lines 3 to 7 show how to compute the semantic similarity ($relevance$) between topic word and post. Line 8 reveals the equation to compute the boosted relevance ($bRelevance$). This equation indicates that if a person tries to spam the system with too many words related to the product, the logarithmic function $bRelevance$ is not increased too much (in our experiment, we set the value of constant factor $\alpha$ to 20). 
Similarly, we compute $bRelevance$ for textual comments based on the relevant words present in it and assign higher weight to the relevant comments and their interactions.  
We drive the graph structure of the group by representing each user of the group as a vertex of the graph and each user interaction such as `like on comment', `like', `comment', `share' as an edge of the graph.  We assign the weights 1, 2, 4 and 8 for like on comment, like, comment, and share respectively~\cite{bucher2012want}. We create an edge from user $u_i$ to $u_j$, if the user $u_i$ has reacted to any post or comment that is created by the user $u_j$. The weights of the edges are determined by the product of the weight corresponding to the type of interaction with the boosted relevance.

\section{Finding Influential Users in OSG}
\label{sec:fipsn_WoMm} 
In this section, we describe how to find influential users in OSGs. This is a solution to problem 2.

We use PageRank~\cite{page1999pagerank} algorithm to find the influential users. 
One of the reasons to use PageRank that it considers the importance of each user while finding the influential user unlike other authority measures~\cite{zhang2007expertise,freeman1977set,freeman1978centrality}. 
PageRank was originally developed to rank the web pages for search results. Web pages are connected together by hyperlinks. 
Similarly, we have topic sensitive social interaction graph where users are connected through social interactions. So, we can  apply PageRank algorithm on the social interaction graph to find influential users. We rank the users based on decreasing order of PageRank score, and select the top-k users to be the potential WoM marketers.

\begin{figure}[h]
    \centering
  \includegraphics[width=4.7cm, height=2.2cm]{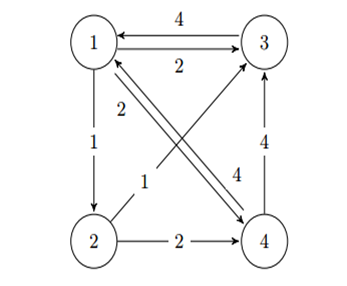}
  \vspace{-0.1in}
    \caption{Social Interaction Graph}
    \label{fig:wpr_sig}
\vspace{-0.1in}
\end{figure}

\noindent \textbf{Example:} 
Consider an example graph in Figure~\ref{fig:wpr_sig}. We apply PageRank on both weighted and unweighted version of the graph to investigate the effect of weighted edges in the computation of users' ranking. 
\vspace{-0.22in}
\vspace{-0.1in}
In this graph, we consider vertices as users and edges as reactions. Weights of each type of reaction are assigned based on type of interaction as described in Section~\ref{sec:fipsn_sig}. We do not consider {$brelevance$} in this example for ease of understanding. 
We apply PageRank on unweighted version of this graph, we get rank of each node as x1 = 0.7210, x2 = 0.2403, x3 = 0.5408, x4 = 0.3605, whereas we get x1=0.7328, x2=0.1466, x3=0.5374, x4=0.3908 for weighted version of the graph. We can see that assigning weights to the edges alters the ranks from the ranks calculated without weights. We can see that x2 has reduced from 0.2403 to 0.1466 because it has only one like on its comment which carries less weight.  
Here, we can also see that x1 has higher rank compare to x3 so it is not only the number reactions determine the authority, it also depends on the importance of users who react on the content. When we apply the PageRank on larger graph, users' rank change significantly as well as their ranking orders. 
We also compare the effectiveness of PageRank algorithm with other authority measure algorithms such as HITS~\cite{kleinberg1999web}, Z-score~\cite{zhang2007expertise}, Eigen vector~\cite{ruhnau2000eigenvector}, Betweenness~\cite{freeman1977set} and Closeness~\cite{freeman1978centrality}.

\section{Reinforced Marketing}
\label{sec:fipsn_rm}
In this section, we give a solution to problem 3. 
In OSGs users interact with other users having similar topics of interests. This type of user interaction 
leads to formation of sub-groups. Since user interactions may get confined to sub-groups, it is important
to find multiple influential users from each sub-group so that they can collectively promote the product, which will 
be more effective in giving trust to users about the product. We need to do this for all the important sub-groups.

We find sub-groups by finding weakly connected components in the graph. 
A weakly connected component is a maximal sub-graph of a directed graph 
such that for every pair of vertices in the sub-graph, 
there is an undirected path. 
So each member of a weakly connected component 
may have reacted on someone’s post in the group or 
would have  received a reaction from someone else in the group. 
We choose to target a sub-group only if it contains enough users. 
If users in the sub-group are less than threshold $th$, we do not select that sub-group for marketing. 

We apply the best authority measure algorithm (described in Section~\ref{sec:fipsn_evl}) 
in the topic sensitive social interaction graph to find the top-$k$ topical influential users of the group.
For each of the sub-group, we select top-$r$ ($r$<$k$) users from the set of $k$ users such that these $r$ users also belong to the sub-group. 
These $r$ influential users can support each other by advertising the same product to their sub-group(s).

\section{Evaluations}
\label{sec:fipsn_evl}
In this section, we describe our dataset and evaluation metrics. We also compare the performance of various authority measure algorithms and show the characteristics of influential users through some anecdotal examples. 

\subsection{Experimental Setup}
We use dataset of Facebook groups for the experiment as these are focused groups with large number of audience. Facebook groups are community of people where they share their common interests in the form of posts and comments. Members of groups can react to the posts/comments created by each other. Reactions consist of a textual comment and a unary rating score in the form of likes and shares. We use Facebook Graph API to collect the dataset. The dataset contains 100 of Facebook groups having at least 20,000 members. It includes 0.3 million posts and 10 million of reactions that were created in 5 years (from 2011 to 2015). 
We perform various text pre-processing tasks on text content of dataset such as stop words removal, stemming and lemmatization.

\subsection{Evaluation Metrics}
We show the effectiveness of algorithms by using three metrics namely correlation, precision, and influence.

{\bf Correlation}: We use correlation metric to measure the strength of association between two ranks. We use Pearson correlation~\cite{lawrence1989concordance} to evaluate the effectiveness of authority finding algorithms. 
We find the correlation between rank assigned to users by authority measure algorithms and the baseline influence metrics (described later in this section). 

{\bf Precision and Normalized Discounted Cumulative Gain}: We use these measures to check the quality of authority finding algorithms by measuring the relevancy of top-k influential users generated by these algorithms. Precision is fraction of retrieved instances that are relevant. Normalized Discounted Cumulative Gain is computed based on the discounted cumulative gain~\cite{jarvelin2002cumulated} which includes the position of users in the consideration of their importance.

{\bf Influence}: We use two influence metrics as baselines to evaluate the user's authority position in the group namely, centrality and popularity. Degree is a centrality measure that evaluates the user's connectivity whereas votes and topical votes are popularity measures that evaluate the user's prestige. For each user, we compute votes by taking the weighted sum of all the audience reactions received by the user over all his posts, comments. However, we compute topical votes by taking the weighted sum of audience reactions over all his posts, comments that contain the advertisement topic itself or the topics semantically related to the advertisement topic. 

\subsection{Effectiveness of Algorithms}
\label{sec:corr}
It is important to understand the effectiveness of authority finding algorithms in OSGs. The algorithm which is appropriate for one network may not be appropriate for other because of user behavior dynamics. To measure the effectiveness of authority finding algorithms in OSGs, we find the correlation of top-200 influential users generated by authority finding algorithms with the votes received by these users. As all of these groups are technical groups having similar number of users, we compute the overall correlation by averaging the correlation across all the groups. 

 \vspace{-0.25in}
\begin{figure}
\centering
\begin{minipage}{.48\columnwidth}
  \centering
  \includegraphics[width=5.5cm, height=4cm]{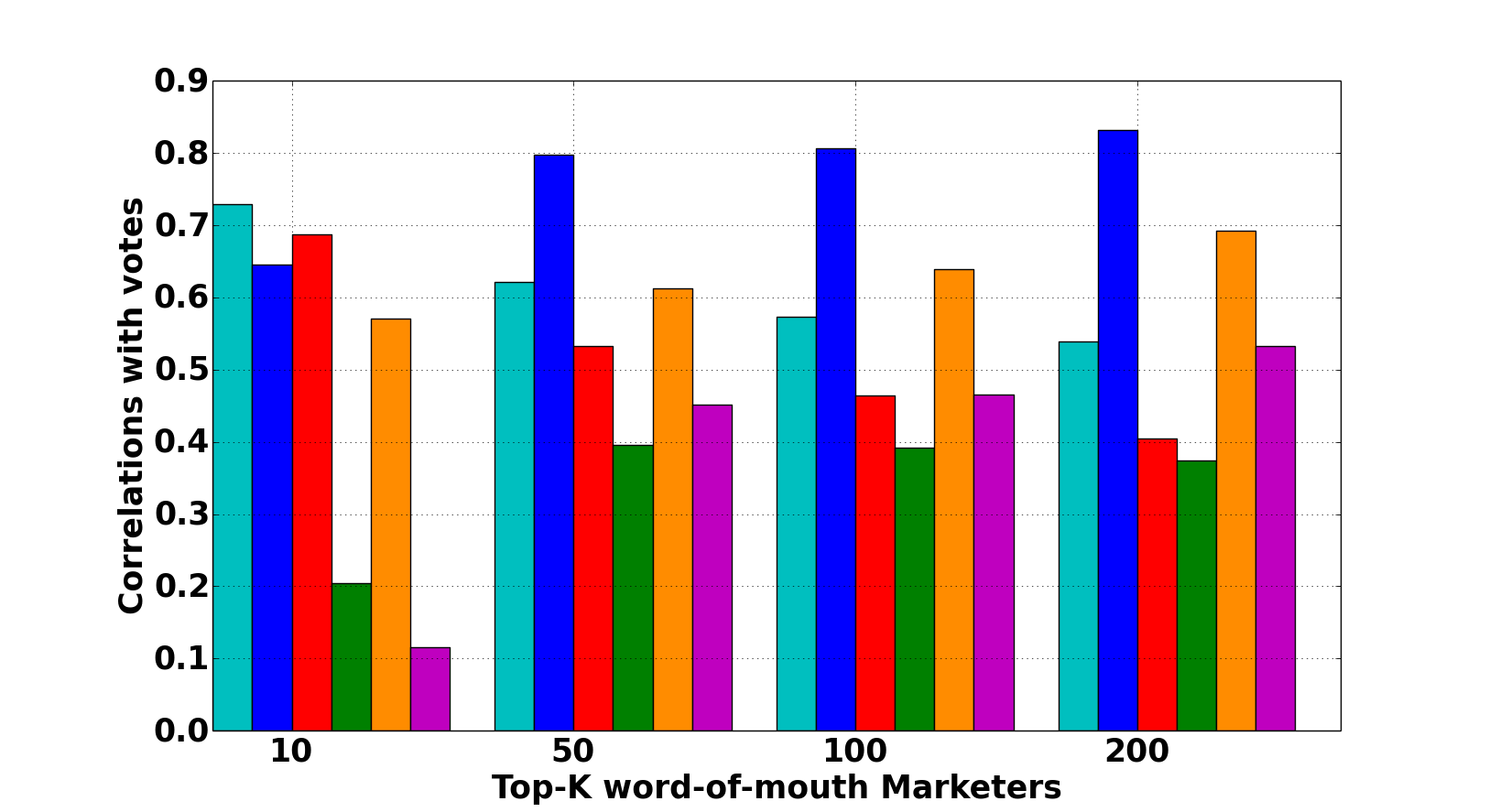}
  \captionof{figure}{Correlation of authority \\ finding algorithms with votes}
  \label{fig:corr1}
\end{minipage}%
\begin{minipage}{.55\columnwidth}
  \centering
  \includegraphics[width=6.5cm, height=4cm]{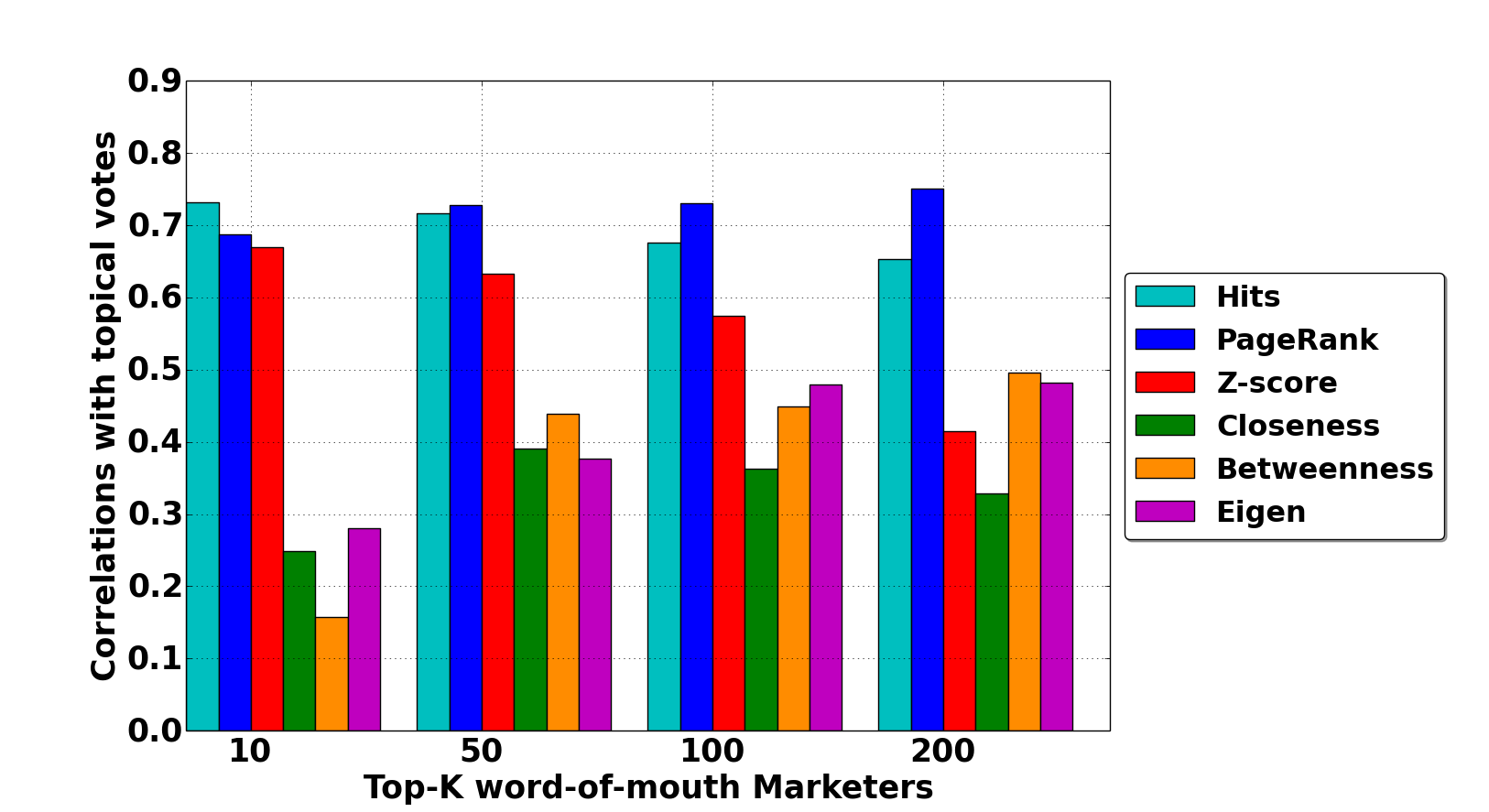}
  \captionof{figure}{Correlation of authority \\ finding algorithms with topical votes}
  \label{fig:corr2}
\end{minipage}
\end{figure}
 \vspace{-0.2in}

Figure~\ref{fig:corr1} shows the correlation of the top-200 influential users ranked by the various authority finding algorithms with the votes. HITS performs better than other algorithms for top-10 users whereas PageRank outperforms for top-50 or more users. One of the reasons is that PageRank is a global measure and it does not trap in local neighborhood however HITS suffers from topic drift.  
Betweenness tends to produce slightly better results than most of the other algorithms. One of the reasons is that nodes having high Betweenness are the bridges of two parts of the graph (sub-graph) and have the potential to disconnect graph if removed. If a user having high Betweenness posts an update, there is high chance that it will spread rapidly across the sub-graphs. 

Figure~\ref{fig:corr2} shows the correlation of the top-200 influential users ranked by the various authority finding algorithms with the topical votes. 
HITS performs better for top-10 users whereas PageRank performs better than HITS for top-50 or more users. 
In conclusion, PageRank can be utilized for finding influential users for general marketing as it shows high correlation with both votes and topical votes. 

\subsection{Precision Analysis}
\label{sec:precision}
We evaluate the correctness of authority finding algorithms by using Mean Average Precision~(MAP) and Normalized Discounted Cumulative Gain~(NDCG), which are standard measures to evaluate the effectiveness of web page ranking algorithms. 
We consider top-50 influential users of the groups generated by algorithms. 
We ask five students of our research lab to join these technical groups and manually judge whether a user is influential or not from their viewpoints for a given topic \textit{T}. 
We also ask to rank these users for a given topic. 
We provide all the posts and reactions of influential users to the students. 
These students label the data independently, without influencing each other. The average percentage of agreement among the students was 92\%. We use this label data as a ground truth for finding the MAP and NDCG of algorithms. We compute the overall MAP, NDCG by averaging the MAP, NDCG across all the groups respectively. 
\vspace{-0.27in}
\begin{table}[!h]
\begin{center}
\small
\begin{tabular}{|c|l|l|}
\hline
\hline
\thead{\bf Authority Measures} & \thead{\bf MAP} & \thead{\bf NDCG}\\
 \hline\hline
 PageRank & 0.91 & 0.83 \\
 \hline
 HITS & 0.87 & 0.75 \\
 \hline
 Z-score & 0.70 & 0.65 \\
 \hline
 Eigen & 0.72 & 0.69 \\
 \hline
 Betweenness & 0.76 & 0.70 \\
 \hline
 Closeness & 0.73 & 0.67 \\
 \hline 
\end{tabular}
\caption{MAP and NDCG of authority finding algorithms}
\label{table:presision}
\end{center}
\end{table}
\vspace{-0.54in}

As can be observed in Table~\ref{table:presision}, for a given topic \textit{T} PageRank performs better than other authority finding algorithms. PageRank finds topic sensitive influential users with the highest accuracy. So, we use PageRank for our analysis in rest of the paper. 

\subsection{Marketing Across Topics}
In this section, we analyze behaviour of influential users across different topics and investigate how widely the rank correlation of these users changes by changing the topics. 

Top influential users (top users) for all the query topics are not different. Top users tend to express their opinions on many popular topics of the group. To examine dynamic behavior of top users across different topics, we compare the relative order of their ranks across topics. We ignore the least popular topics and focus on the set of relatively popular topics. 
We apply Topical N-Grams~\cite{wang2007topical} on the posts to find popular topics of the Java For Developers\footnote{https://www.facebook.com/groups/java4developers/} group (Java group). Web, Servlet, and Constructor are some popular topics in the group, so we choose these topics to measure the variation in top users ranking across these topics. We use correlation to compare the ranking patterns of top users for pairs of topics.

\vspace{-0.3in}
\begin{table}[ht]
\begin{minipage}[b]{0.46\linewidth}
\centering
\small
  \begin{tabular}{|l|l|l|}
\hline
\hline
\thead{\bf Topics} & \thead{\bf Top-20 \\ \bf Users} & \thead{\bf Top-200 \\ \bf Users}\\
 \hline\hline
 Web vs. Servlet & 0.79 & 0.56 \\
 \hline
 Web vs. Constructor & 0.53 & 0.46 \\
 \hline
 Constructor vs. Servlet & 0.49 & 0.39 \\
 \hline
\end{tabular}
    \caption{Correlation in top users ranking for popular topics}
    \label{table:infactopics}
\end{minipage}\hfill
\begin{minipage}[b]{0.45\linewidth}
\centering
\includegraphics[width=5.5cm, height=3cm]{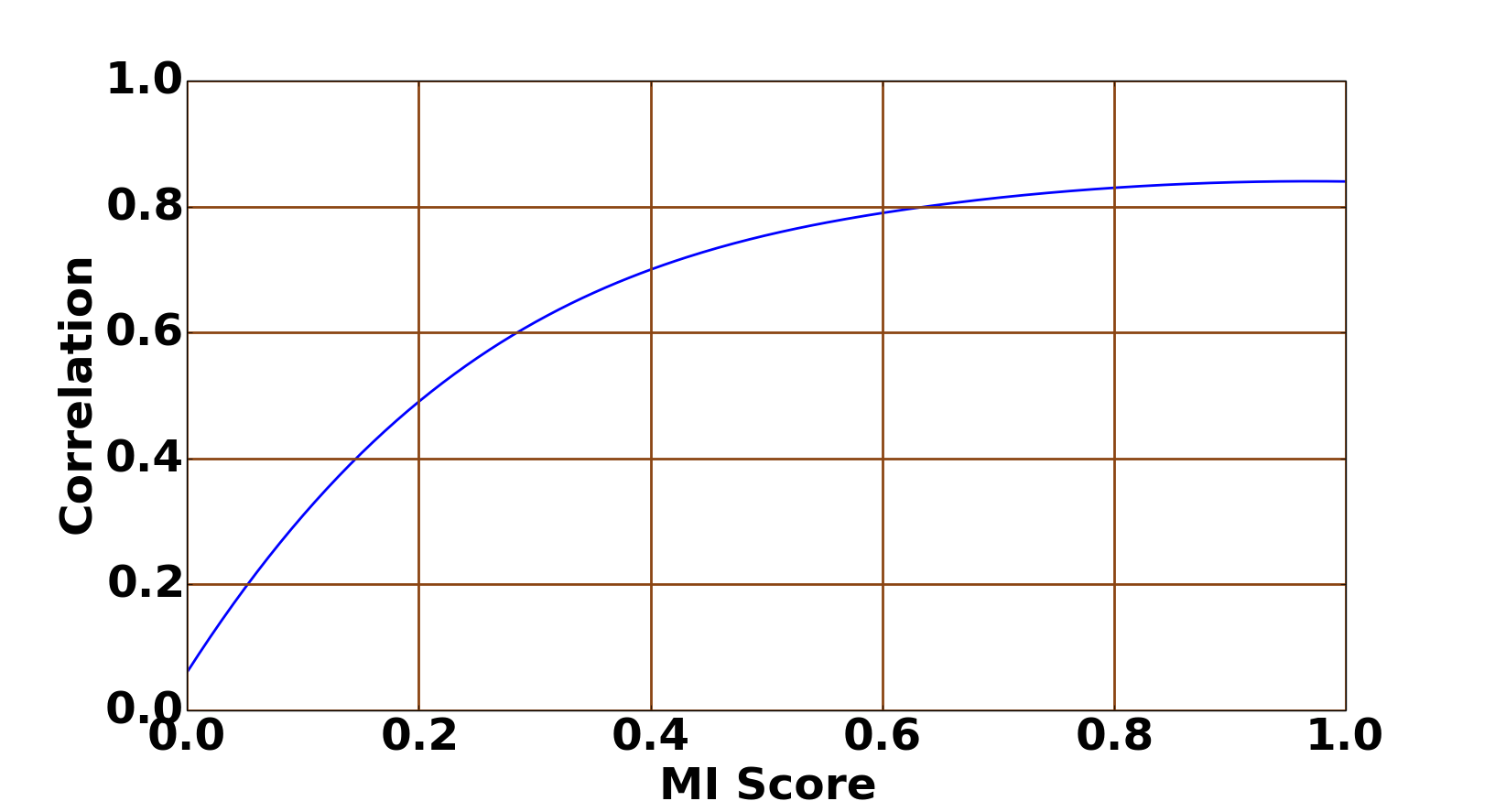}
\captionof{figure}{Correlation of top users \\ across variety of topics}
\vspace{-0.01in}
\label{fig:mivscorr}
\end{minipage}
\end{table}
\vspace{-0.3in}

We observe in Table~\ref{table:infactopics} that correlation is high for the top-20 users  which implies that these users post over a wide range of topics. Among topic pairs, \{Web, Servlet\} shows the highest correlation for the top-20 users. This is because these two topics are closely related in Java. 
Servlets are used in Web programming. 
This analysis indicates that top users hold significant influence over a range of topics and could be used to spread the information about variety of topics. 

To get more insight into variation in correlation of top users across topics, we perform the experiment on wide range topics in Java group. We select 20 topics from each of popular topics, less popular topics and unpopular topics. We compute Mutual Information~(MI) score for all these topics with respect to group topic (shared group interest). 
We derive top-200 topic sensitive influential users by using these topics and measure the correlation of these users with topical votes. 
As can be observed in Figure~\ref{fig:mivscorr} that correlation decreases as MI score decreases. If a chosen topic has very less dependency with the group topic, then authority measure algorithms show very less correlation. 
It indicates that quality of top users also depends on the topic. If a query topic is less related to shared group interest then it is not possible to get prominent topical users who can influence the whole group as the quality of these users decreases. Therefore, it is recommended that advertising business should select a query topic which is highly related to shared group interest to do effective marketing in OSGs.

\subsection{Empirical Evaluation}
In order to investigate influential users' characteristics and behavior dynamics, we find the connectivity of influential users and their structural position in OSGs. First, we find indegree connectivity of top-k influential users in the Java For Developers group (Java group) having 35,000 members at the time of experiment. 
We observe that average indegree of top-20 users is 1604 whereas average indegree of the whole group is 8. The reason for this is that authority finding algorithm strongly correlate with the indegree of the top users. 
Moreover, we observe that 6.5\% users post the 85\% content of the group content and less than 2\% of them are able to influence 80\% users of the group. 

To get more insight into the structural position of influential users in the group, we present the network structure of influential users which is a undirected network constructed in a similar way as mentioned in Section~\ref{sec:fipsn_sig}.
We take a small instance of Java group with 707 nodes, 1187 edges and visualize the network structure of the Group. The users of the network can be divided into two types: top users and ordinary users. The green color nodes represent the top-20 users, and the red color nodes represent ordinary users of the group. 
 \vspace{-0.12in}
\begin{figure}[h]
    \centering
   \fbox{\includegraphics[width=8cm, height=3.5cm]{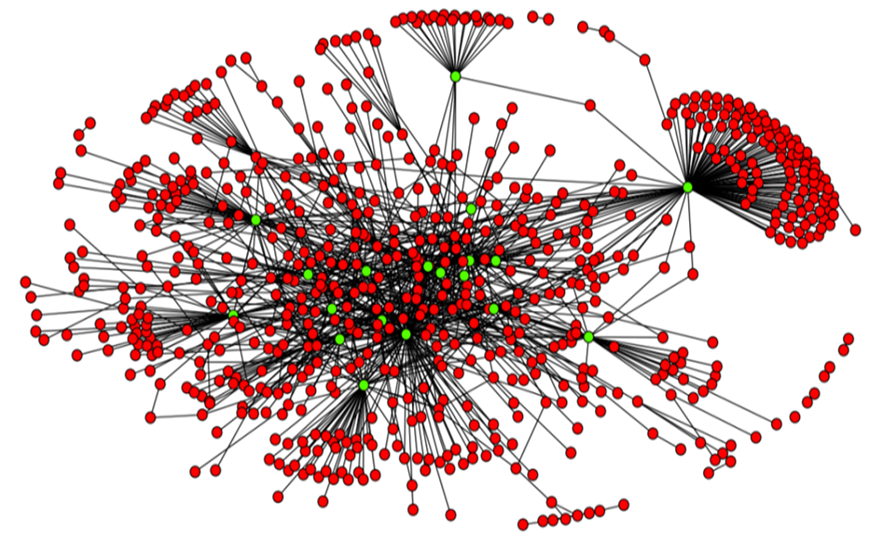}}
    \caption{Structure of Java group}
    \label{fig:java_for_dev}
     \vspace{-0.21in}
\end{figure}

Figure~\ref{fig:java_for_dev} shows that top users are strongly connected with the large number of members of group. 
Statistics reveal that average degree of the group is 3.35 whereas average degree of top-20 influential users is 72. 
Moreover, average number of reactions received by a user of the group is 5.2 whereas average reaction received by top-20 influential users is 98. 

Furthermore, we also analyze the reactions received by users in the Java group to examine the popularity of influential users in the group. Top users receive large number of reactions as these are the prestigious users of the group. As rank of the user increases, reaction received by the user decreases exponentially. This difference indicates that it is more beneficial to target popular users for marketing than to employ a massive number of non-popular users. 

\subsection{Temporal Dynamics} 
We analyze the action (posting) and reaction behavior of influential users over a period of time and find the right time to start promotion in the group to maximize content visibility. Our results are based on 5 years of temporal data.  

\begin{figure}
\centering
\begin{minipage}{.52\columnwidth}
  \centering
  \includegraphics[width=6.3cm, height=3cm]{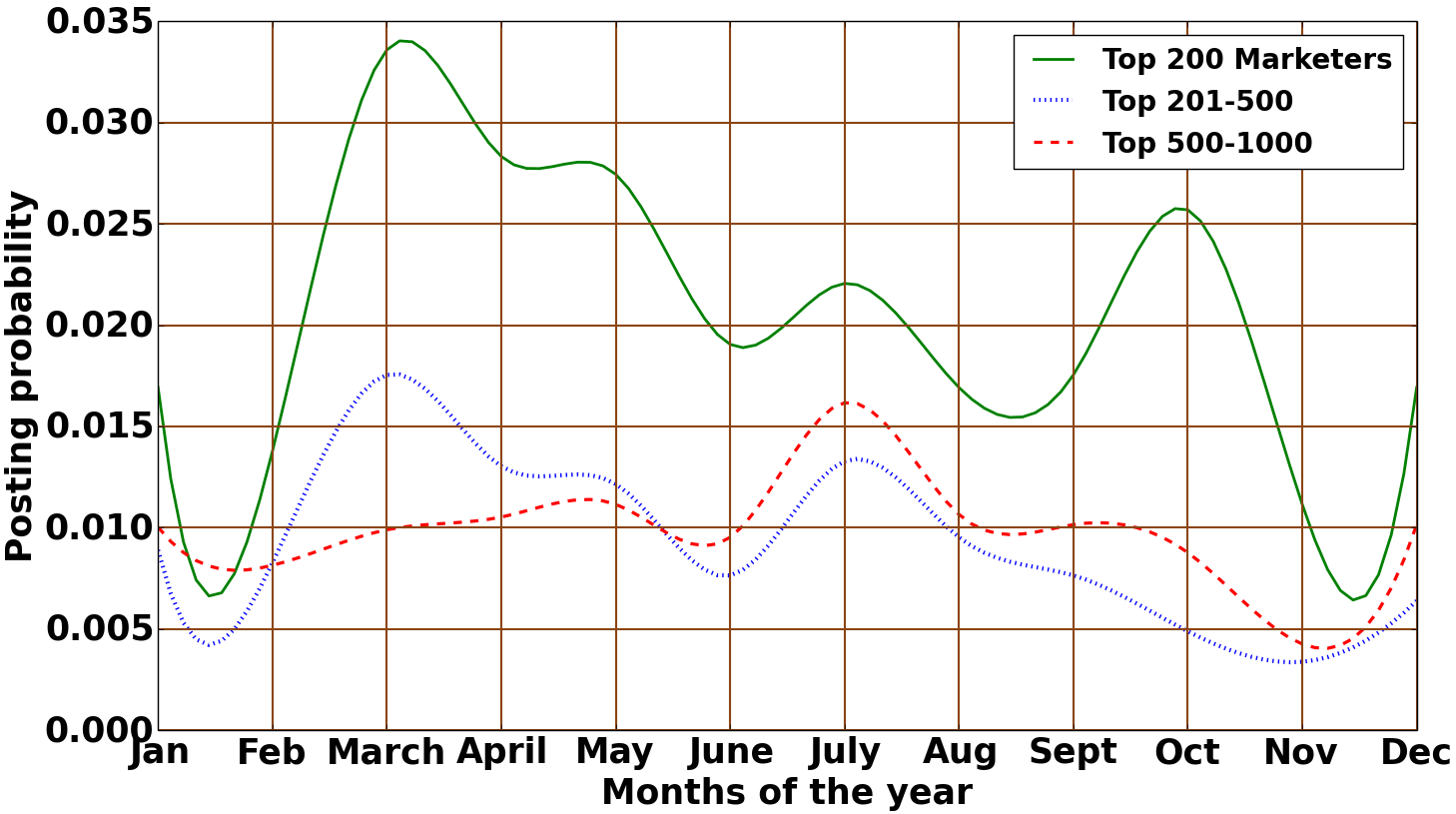}
  \captionof{figure}{Posting behavior of top users}
  \label{fig:posting}
\end{minipage}%
\begin{minipage}{.5\columnwidth}
  \centering
  \includegraphics[width=6.0cm, height=3cm]{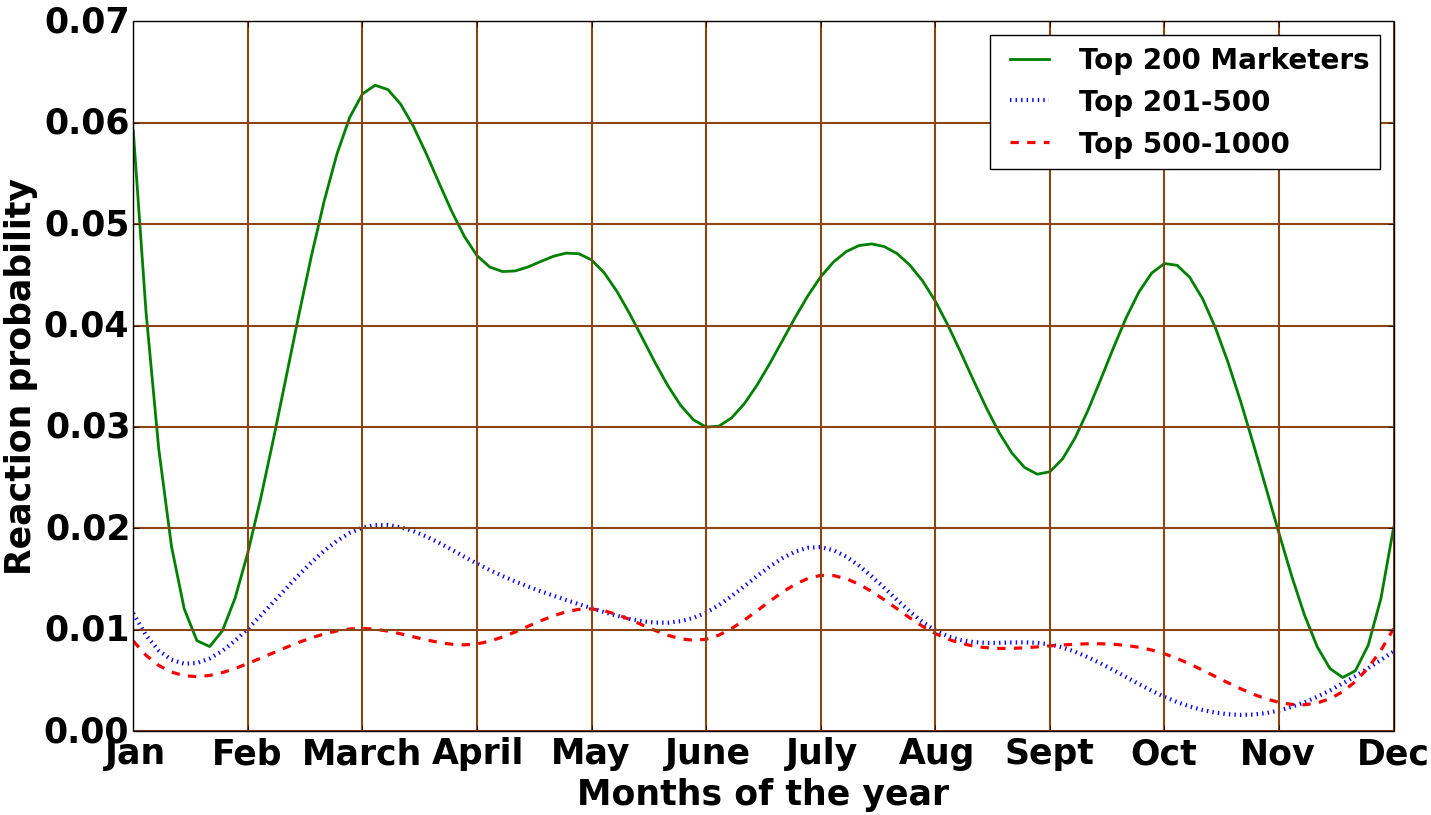}
  \captionof{figure}{Reaction behavior of top users}
  \label{fig:reaction}
\end{minipage}
\end{figure}

In order to examine the influential users' posting behavior, we pick the \linebreak top 1000 influential users based on their ranks from the Java group. We divide top users into three groups based on their ranks such as top 200 users,  \linebreak top 201-500 users and top 501-1000 users. Our aim is to analyze the differences in posting behavior of these users. We compute the probability of posting a post for all these three groups in each month of the year. 
Figure~\ref{fig:posting} shows the time evolution of the posts of the influential users (top users).

Our findings about the posting behavior of top users reveal two interesting observations. First, top users post significant updates over a period of time.  \linebreak Top 200 users post lots of information compared to top 500 and top 1000 users. 
Second, lots of posts are posted during the month of March, April, and October. This is perhaps due to various competitive and semester exams in India during these months, which motivates the top users to post a lot of information about various topics. So, it is better to choose these periods of the year for marketing. 
 \vspace{-0.19in}
 \vspace{-0.21in}

We also perform the similar experiment on reactions received by top users. As can be seen in Figure~\ref{fig:reaction}, reaction pattern follows the same trend as posting pattern, i.e., more number of audience reactions in the month of March, April, and October. It is due to a large number of posts created by top users during these periods of months and this posting behavior leads to increase the number of audience reactions. As lots of users are active during these periods of months, advertising companies can target more number of top users to promote products during these periods.

\section{Conclusion}
\label{sec:fipsn_cnfw}
In the paper, we propose methods to use OSGs for WoM marketing. 
We present an algorithm to create topic sensitive social interaction graph from the activities of the group. 
We apply authority finding algorithm on social interaction graph to find topic specific influential users. 
Organizations can promote the product through these influential users by giving them incentives. 
We propose the concept of reinforced marketing to perform effective marketing where multiple influential users collectively market a product. We also analyze the important characteristics of influential users such as these users post most of the content of the group and able to influence most of the population of the group. We find that influential users post over a wide range of topics and receive lots of audience reactions. Finally, we show the best time of the year to start marketing in Facebook groups to improve the effectiveness of marketing.

 \bibliography{bibliography} 

\begin{thebibliography}{10}
\providecommand{\url}[1]{\texttt{#1}}
\providecommand{\urlprefix}{URL }

\bibitem{broder2000graph}
Broder, A., Kumar, R., Maghoul, F., Raghavan, P., Rajagopalan, S., Stata, R.,
  Tomkins, A., Wiener, J.: Graph structure in the web. Computer networks
  (2000)

\bibitem{bucher2012want}
Bucher, T.: Want to be on the top? algorithmic power and the threat of
  invisibility on facebook. new media \& society  14(7) (2012)

\bibitem{chen2009efficient}
Chen, W., Wang, Y., Yang, S.: Efficient influence maximization in social
  networks. In: SIGKDD. ACM (2009)

\bibitem{cheng2014can}
Cheng, J., Adamic, L., Dow, P.A., Kleinberg, J.M., Leskovec, J.: Can cascades
  be predicted? In: WWW. ACM (2014)

\bibitem{domingos2001mining}
Domingos, P., Richardson, M.: Mining the network value of customers. In:
  SIGKDD. ACM (2001)

\bibitem{MarketShare}
Forbes: What are they saying about your brand?
  \url{http://www.forbes.com/sites/pauljankowski/2013/02/27/quick-what-are-they-saying-about-your-brand/#ee5ff7374a8d}
  (2013)

\bibitem{freeman1978centrality}
Freeman, C, L.: Centrality in social networks conceptual clarification. Social
  networks  (1978)

\bibitem{freeman1977set}
Freeman, L.C.: A set of measures of centrality based on betweenness. Sociometry
   (1977)

\bibitem{guille2013information}
Guille, A., Hacid, H., Favre, C., Zighed, D.A.: Information diffusion in online
  social networks: A survey. SIGMOD  42(2) (2013)

\bibitem{incite}
Incite: How social media amplifies the power of word-of-mouth.
  \url{http://www.incite-group.com/brand-management/how-social-media-amplifies-power-word-mouth}
  (2014)

\bibitem{jarvelin2002cumulated}
J{\"a}rvelin, K., Kek{\"a}l{\"a}inen, J.: Cumulated gain-based evaluation of ir
  techniques. ACM Transactions on Information Systems (TOIS)  20(4) (2002)

\bibitem{kempe2003maximizing}
Kempe, D., Kleinberg, J., Tardos, {\'E}.: Maximizing the spread of influence
  through a social network. In: SIGKDD. ACM (2003)

\bibitem{kleinberg1999web}
Kleinberg, J.M., Kumar, R., Raghavan, P., Rajagopalan, S., Tomkins, A.S.: The
  web as a graph: measurements, models, and methods. In: International
  Computing and Combinatorics Conference. Springer (1999)

\bibitem{lawrence1989concordance}
Lawrence, I., Lin, K.: A concordance correlation coefficient to evaluate
  reproducibility. Biometrics  (1989)

\bibitem{leskovec2007patterns}
Leskovec, J., McGlohon, M., Faloutsos, C., Glance, N.S., Hurst, M.: Patterns of
  cascading behavior in large blog graphs. In: SDM. SIAM (2007)

\bibitem{page1999pagerank}
Page, L., Brin, S., Motwani, R., Winograd, T.: The pagerank citation ranking:
  bringing order to the web.  (1999)

\bibitem{ruhnau2000eigenvector}
Ruhnau, B.: Eigenvector-centrality—a node-centrality? Social networks  (2000)

\bibitem{trusov2010determining}
Trusov, M., Bodapati, A.V., Bucklin, R.E.: Determining influential users in
  internet social networks. Journal of Marketing Research  47(4) (2010)

\bibitem{vogiatzis2013influential}
Vogiatzis, D.: Influential users in social networks. In: Semantic
  Hyper/Multimedia Adaptation. Springer (2013)

\bibitem{wang2007topical}
Wang, X., McCallum, A., Wei, X.: Topical n-grams: Phrase and topic discovery,
  with an application to information retrieval. In: Data Mining. ICDM 2007

\bibitem{weng2010twitterrank}
Weng, J., Lim, E.P., Jiang, J., He, Q.: Twitterrank: finding topic-sensitive
  influential twitterers. In: WSDM. ACM (2010)

\bibitem{wu2011says}
Wu, S., Hofman, J.M., Mason, W.A., Watts, D.J.: Who says what to whom on
  twitter. In: WWW. ACM (2011)

\bibitem{zhang2007expertise}
Zhang, J., Ackerman, M.S., Adamic, L.: Expertise networks in online
  communities: structure and algorithms. In: WWW. ACM (2007)

\end{thebibliography}
 \bibliographystyle{splncs03}

\end{document}